\documentclass[12pt]{article} 
\usepackage[french]{babel} 
\usepackage[T1]{fontenc} 
\usepackage[utf8]{inputenc}
\begin{document}  

\begin{center}
{\bf  Young Gerard storming high energy physics}
\vskip 1cm
A personal recollection\footnote{To appear in a special issue of {\it Int. Journal of Mod. Phys.} A, Vol. 41, No 26.}
\vskip 1.5 cm
John Iliopoulos
\vskip 0.5cm
{\it 
 Laboratoire de Physique de l'Ecole Normale Supérieure\\
 ENS-PSL, CNRS, Sorbonne Univ., Univ. Paris Cité, Paris, France} \\
(jean.iliopoulos@phys.ens.psl.eu)
\end{center}
\vskip 2cm
{\bf Abstract :} On the occasion of Gerard ’t Hooft’s 80th birthday, I present my personal recollections of a scientific meeting held in the Orsay Campus of the University of Paris in January 1972. In this meeting Gerard gave for the first time a detailed account of his groundbreaking work on the renormalization properties of Yang-Mills theories, both in the massless phase as well as in the phase with spontaneous symmetry breaking.
\vskip 2cm

Gerard 't Hooft made a spectacular entrance in the world of theoretical high energy physics with two papers, both published in 1971: “Renormalization of massless Yang-Mills fields”\cite{Hooft1} and “Renormalizable Lagrangians for massive Yang-Mills fields”\cite{Hooft2}. They were part of his PhD thesis. Rarely a {\it coup d'essai} has ever been such a {\it coup de ma\^{i}tre}. And even more rare is the example of a young student's thesis having such a groundbreaking and lasting impact on physics. I remember Steven Weinberg saying in a talk about  the history of this period “\dots and then, all hell broke loose”\footnote{A measure of this impact:  Weinberg's 1967 paper which presents the Standard Model for leptons, went totally unnoticed when it was first published. It has practically no citations until 1972. Even Weinberg himself seemed to have forgotten it. It is only after 1972, that is after the publication of 't Hooft's papers, that the number of citations skyrocketed. I have presented this part of the story in references \cite{Ilio1} and \cite{Ilio2}.}.

All this is well known and amply documented, but in this short note I want to tell a story which is part of the big picture but has remained rather confidential. Eye-witnesses were few when it happened and, as years go by, even fewer can still testify about it. I happened to be the initiator of this event and I will present here my recollection of it, although I am fully aware that memories have turned dim.

My own interest in the quantum theory of Yang-Mills fields was rather late. In my early work on the weak interaction divergence structure I was assuming the existence of massive intermediate vector bosons but with no Yang-Mills self couplings. The calculations in our paper on the charmed quark with Sheldon Glashow and Luciano Maiani were done in that framework. However, in the version we submitted to the Physical Review, we added a remark saying that this theory could be extended to a full Yang-Mills, leaving aside the question of the vector boson masses. The answer came immediately in the form of a referee's report. He (or she) had read our paper very carefully, said it was interesting and worth publishing, but raised one objection. For the Intermediate Vector Boson theory, the $n$th order diagrams give a leading divergence of $(G\Lambda^2)^n$. In our paper we were implicitly assuming that the same remains true even in the presence of the Yang-Mills interaction. However, as the referee pointed out, a na\"ive power counting for massive Yang-Mills bosons gives instead $(G\Lambda^6)^n$. He then went on to remark: “This behavior can undoubtedly be improved, but the assertion that it can go down to $(G\Lambda^2)^n$ must be either proven or deleted.” He did not say it was wrong, which proves that he knew the problem very well. We immediately realized that proving the $\Lambda^{2n}$  behavior was not easy, so we decided to change one or two sentences and resubmit. This time the paper was accepted. However, the question triggered our curiosity and during the following months Glashow and myself -- Luciano Maiani had already left for Rome -- devoted quite some time to answer it. We succeeded and we published two papers on it\cite{Gl-Il}, both appearing in 1971. I still consider them as the most intelligent papers I have ever written, although they have hardly been read by anyone. They were totally eclipsed by the two 't Hooft papers. I had tried to go one step further in the divergence cancellation and prove renormalizability, but I failed. As a result, by the fall of 1971, sort of accidentally, I found myself  interested in the renormalization properties of Yang-Mills theories.  When I returned to Paris I heard that a young student of Tini Veltman had solved the problem. I wanted to learn the details and decided to organize a meeting gathering people who could be interested. They were not very many, the subject was not yet very popular. 

The meeting took place at the University of Paris in Orsay, I believe in January 1972, I do not remember the exact dates. The participants were the local people from Orsay, Claude Bouchiat, Philippe Meyer et al, some theorists from the nearby Saclay Center and, of course, Tini Veltman and a few among his students from Utrecht. Obviously, Gerard was among them. I had also invited some other theorists mainly from CERN. I remember, in particular, William Bardeen, Henri Epstein, Raymond Stora, Bruno Zumino and probably a few others. 

The meeting started as a normal informal workshop. There were a few talks, typically one hour long, with many questions and remarks. Participants were encouraged to interrupt the speaker and ask questions during the talks. I remember I gave a talk on the cancellation of the axial anomalies between leptons and quarks in the $SU(2)\times U(1)$ model. I think it was Veltman who presented the dimensional regularization scheme. And then Gerard was asked to present his work on the renormalization properties of Yang-Mills. I assume that he had prepared a one-hour talk, like everybody else. But as soon as he started talking, the meeting 
underwent a sudden phase transition. I had set the rule that I would not interrupt an interesting discussion in order to keep any kind of schedule. And discussions there were! We thought we knew quantum field theory, but in fact, for most of us this knowledge was limited to the perturbation expansion of the simplest $\phi^4$ model. All the subtle questions related to the quantization of constrained systems, such as Yang-Mills theories, were new to many of us. And here there was this young man who seemed to know everything. He was the youngest participant in the meeting but it became soon obvious that he was the one who was running the show! Unperturbed, showing no signs of fatigue or confusion, he kept on answering in great detail all sorts of questions. Before entering the real problem of the renormalizability, he had to explain to us many “technical” questions:  the various gauge choices -- renormalizable gauges, unitary gauges, the gauge which became known as “the 't Hooft gauge”-- the way to obtain the Feynman rules, including the Faddeev-Popov ghosts, for each one of them, the generalized Ward identities etc. He had reformulated the Brout-Englert-Higgs mechanism and could prove the gauge invariance and the unitarity of the resulting $S$-matrix. Time went on and late in the afternoon we decided to stop and continue next morning. At 9 am Gerard was again in front of the blackboard continuing the presentation from the point he had left it the previous day.  And the same story was repeated. Late in the afternoon we stopped again to reconvene the following day and so on. I believe he talked altogether for something like 15 or 18 hours. By the end of the week the audience was exhausted, but Gerard seemed to be time translational invariant. I am sure he could have kept the same pace for a month.  

The discussions themselves were lively and very interesting and the audience often took an active part. I remember, in particular, the discussion which helped clarifying the  calculation of the axial anomaly in the dimensional regularization scheme. Bill Bardeen played an important role in it. Raymond Stora told me later that it was after this meeting that he decided to deepen the meaning of gauge invariance. The discovery of the BRS symmetry by the Marseille group was a direct consequence. 

As it is often the case with successful meetings, this one had several descendants. The University of Marseille organized every summer a similar meeting for several years. The Universities of Paris, Rome and Utrecht initiated a series, which became known as “triangular meetings”. Subsequently enlarged with the participation of other European labs, they played an important role in federating the European community of theoretical high energy physics. They gave rise to European Research Networks which were funded later by the European Community. In fact, scientists built the scientific European union before politicians thought of the economic one. 

The success of this Orsay meeting was largely due to Gerard 't Hooft. It was probably his first public appearance, certainly the first in which he played the central role. All participants to this meeting left with the clear feeling that a star was born. This is a fading and personal recollection.

\end{document}